\documentstyle[11pt]{article}
\textwidth\def\kon#1#2{\vbox{\halign{##&&##\cr\lower4pt
\hbox{$\scriptscriptstyle\vert$}\hrulefill &\hrulefill\lower4pt
\hbox{$\scriptscriptstyle\vert$}\cr $#1$&$#2$\cr}}}

\def\fii{\varphi}
\def\al{\alpha}
\def\be{\beta}
\def\ga{\gamma}
\def\ro{\varrho}

\def\eh{{\scriptstyle{1\over 2}}}
\def\d{\partial}
\def\=d{\,{\buildrel\rm def\over =}\,}

\def\sqr#1#2{{\vcenter{\vbox{\hrule height.#2pt\hbox{\vrule width.
#2pt height#1pt \kern#1pt \vrule width.#2pt}\hrule height.#2pt}}}}

\def\la{\lambda}

\def\sq{\hbox{\rlap{$\sqcap$}$\sqcup$}}

\def\te{\vartheta}

\begin{document}

\title{ FROM MASSIVE GRAVITY TO DARK MATTER DENSITY}
\author{G. Scharf
\footnote{e-mail: scharf@physik.unizh.ch}
\\ Institut f\"ur Theoretische Physik, 
\\ Universit\"at Z\"urich, 
\\ Winterthurerstr. 190 , CH-8057 Z\"urich, Switzerland}

\date{}

\maketitle\vskip 3cm

\begin{abstract} Massive gravity previously constructed as the
spin-2 quantum gauge theory is studied in the classical limit.
The vector graviton field $v$ which does not decouple in the limit
of vanishing graviton mass gives rise to a modification of
general relativity. The modified Schwarzschild solution contains
a contribution from the $v$-field which can be interpreted as the
dark matter mass density. We calculate the corresponding density 
profile in the simplest spherical symmetric geometry.

\end{abstract}

\newpage

\section{Introduction}

Massive gravity is a controversal subject because it is very difficult
to construct it starting from a classical lagrangian theory. On the other
hand, there exists a powerful method to find the quantum theory directly:
the derivation of gauge couplings from a cohomological formulation of
gauge invariance in terms of asymptotic free fields. This method works
for massless and massive gauge theories equally well. For spin-1 theories
this was demonstrated in the monograph [1], where the massless spin-2
case is also treated. The massive spin-2 theory was first investigated
in [2]. The most elegant way to obtain the theory is by assuming the
gauge invariance condition for all chronological products in the form of
the descent equations [3]. These give the total interaction Lagrangian
including ghost couplings and the necessary coupling to a vector-graviton
field. By comparison with the couplings derived from expansion of the
Einstein Lagrangian in powers of the coupling constant we observe that
we really have a quantum theory of gravity. The massive spin-2 theory
corresponds to the Einstein Lagrangian with a $negative$ cosmological
constant. So this is not directly related to the dark energy (contrary
to our earlier belief).

By analogy with spin-1 one would expect that if
the graviton is massive, a gravitational Higgs field would be
necessary to save gauge invariance in higher orders. This would
be a natural candidate for dark matter.
However, the theory
is gauge invariant in second order without any additional field [2], and
in third order the anomalies checked so far also cancel. A gravitational
Higgs seems not to exist.

There is a second option in massive gravity. A massive spin-2
particle has 5 degrees of freedom in contrast to 2 of the massless
graviton. It turns out that there are even 6 physical modes in massive
gravity [4]. In order to have a smooth massless limit $m\to 0$ of the
massive theory, one is forced to choose the 6 physical states as follows: 
two of them are the transversal modes of the massless graviton, the remaining
four are generated by a vector field $v^\lambda$ (called vector-graviton
field, see next section) which is required in
the massive theory to have it gauge invariant. The surprising fact is
that for $m\to 0$ this vector-graviton field $v^\lambda$ does not
decouple from the symmetric tensor field $h^{\mu\nu}$ which corresponds
to the classical $g^{\mu\nu}$ of Einstein. Consequently, the massless
limit of massive gravity is not general relativity; there is a
modification due to the (now massless) vector graviton field
$v^\lambda$. We will show that this modification gives rise to an
additional force which looks as if it comes from a dark matter density.

\section{Massive quantum gravity}

The basic free asymptotic fields of massive gravity are the symmetric
tensor field $h^{\mu\nu}(x)$ with arbitrary trace, the fermionic ghost
$u^\mu(x)$ and anti-ghost $\tilde u^\mu(x)$ fields and the
vector-graviton field $v^\lambda(x)$. They all satisfy the Klein-Gordon 
equation
$$(\sq+m^2)h^{\mu\nu}=0=(\sq+m^2)u^\mu=(\sq+m^2)\tilde u^\mu
=(\sq+m^2)v^\la    \eqno(2.1)$$
and are quantized as follows [2] [3]
$$[h^{\alpha\beta}(x), h^{\mu\nu}(y)]=-ib^{\alpha\beta\mu\nu}D_m(x-y)
\eqno(2.2)$$
with
$$b^{\alpha\beta\mu\nu}=\eh(\eta^{\alpha\mu}\eta^{\beta\nu}+\eta^
{\alpha\nu}\eta^{\beta\mu}-\eta^{\alpha\beta}\eta^{\mu\nu}),\eqno(2.3)$$
$$\{u^\mu(x),\tilde u^\nu(y)\}=i\eta^{\mu\nu}D_m(x-y)\eqno(2.4)$$
$$[v^\mu(x),\, v^\nu(y)]={i\over 2}\eta^{\mu\nu}D_{m}(x-y), 
\eqno(2.5)$$
and zero otherwise. Here, $D_m(x)$ is the Jordan-Pauli distribution 
with mass $m$ and $\eta^{\mu\nu}={\rm diag}(1,-1,-1,-1)$ the Minkowski 
tensor.

The gauge structure on these fields is defined through a nilpotent
gauge charge operator $Q$ satisfying
$$Q^2=0, \quad Q\Omega=0\eqno(2.6)$$
where $\Omega$ is the Fock vacuum and
$$d_Q h^{\mu\nu}= [Q,h^{\mu\nu}]=-{i\over 2}(\d^\nu u^\mu+\d^\mu u^\nu
-\eta^{\mu\nu}\d_\al u^\al) \eqno(2.7)$$
$$d_Q u^\mu\=d \{Q,u\}=0$$
$$d_Q\tilde u^\mu\=d \{Q,\tilde u^\mu\} =i(\d_\nu h^{\mu\nu}-m v^\mu)
\eqno(2.8)$$
$$d_Q v^\mu\=d [Q, v^\mu]=-{i\over 2}mu^\mu.\eqno(2.9)$$
The vector-graviton field $v^\la$ is necessary for nilpotency of $Q$. The
vacuum $\Omega$ exists on Minkowski background only, therefore, apart
from simplicity this background is preferred for physical reasons.

The coupling $T(x)$ between these fields follows from the gauge
invariance condition [2]
$$d_Q T(x)=i\d_\al T^\al(x)\eqno(2.10)$$
where $T$ and $T^\al$ are normally ordered polynomials with ghost
number 0 and 1, respectively. In addition we may require the descent
equations 
$$\d_QT^\al=[Q,T^\al]=i\d_\be T^{\al\be}\eqno(2.11)$$
$$[Q,T^{\al\be}]=i\d_\ga T^{\al\be\ga}\eqno(2.12)$$
where the new $T$'s are antisymmetric in the Lorentz indices.
The essentially unique coupling derived from (2.10-12) is given by [3]
$$T=h^{\al\be}\d_\al h\d_\be h-2h^{\al\be}\d_\al h^{\mu\nu}\d_\be 
h_{\mu\nu}-4h_{\al\be}\d_\nu h^{\be\mu}\d_\mu h^{\al\nu}$$
$$-2h^{\al\be}\d_\mu h_{\al\be}\d^\mu h+4h_{\al\be}\d^\nu h^{\al\mu}
\d_\nu h_\mu^\be+4h^{\al\be}\d_\al v_\la\d_\be v^\la$$
$$+4u^\mu\d_\be\tilde u_\nu\d_\mu h^{\nu\be}-4\d_\nu u_\be\d_\mu\tilde u^\be 
h^{\mu\nu}+4\d_\nu u^\nu\d_\mu\tilde u_\be h^{\be\mu}$$
$$-4\d_\nu u^\mu\d_\mu\tilde u_\be h^{\nu\be} 
-4mu^\al\tilde u^\be\d_\al v_\be-m^2\Bigl({4\over 3}h_{\mu\nu} 
h^{\mu\be}h^\nu_\be$$
$$-h^{\mu\be}h_{\mu\be}h+{1\over 6}h^3\Bigl).\eqno(2.13)$$
Here $h=h_\mu^\mu$ is the trace and a coupling constant is arbitrary.
The quartic couplings follow from second order gauge invariance and so on.

We consider the limit $m\to 0$ in the following. The massless limit of
massive gravity is certainly a possible alternative to general relativity.
The new physics comes from the surviving coupling term of the
vector-graviton
$$T_v=h^{\al\be}\d_\al v_\la\d_\be v^\la.\eqno(2.14)$$
Since all fields are now massless we may compare $T_v$ with the
photon-graviton coupling which in linear gravity is given by
$h^{\al\be}t_{\al\be}$, where $t_{\al\be}$ is the energy-momentum tensor
of the electromagnetic field. It is traceless and conserved, $\d^\al
t_{\al\be}=0$. Expressing the field tensor $F^{\mu\nu}$ by the vector
potential $A^\la$ we get a coupling term of the form (2.14) plus four
different additional terms. That means $T_v$ is different from
photon-graviton coupling, and in fact it does not come from a traceless
conserved tensor $t_{\al\be}$. For relativists this may be odd, but we
must emphasize that our first principle is gauge invariance, general
relativity is secondary.

To understand the situation better we consider a general coupling
between $h^{\al\be}$ and a symmetric tensor $t^{\mu\nu}$ with zero gauge
variation $d_Qt^{\mu\nu}=0$:
$$T=h^{\al\be}[a(\eta_{\al\mu}\eta_{\be\nu}+\eta_{\al\nu}\eta_{\be\mu})
+b\eta_{\al\be}\eta_{\mu\nu}]t^{\mu\nu}.\eqno(2.15)$$
From (2.7) we get
$$d_QT=-{i\over 2}(\d^\al u^\be+\d^\be u^\al-\eta^{\al\be}\d_\la u^\la)
(2at_{\al\be}+b\eta_{\al\be}t_\mu^\mu)$$
which by partial integration is equal to
$$={\rm div}+iau^\be\d^\al t_{\al\be}-i(a+b)u^\be\d_\be t_\mu^\mu$$
where div is a divergence $i\d_\al T^\al$. Now gauge invariance (2.10)
requires the vanishing of the additional terms
$$\d^\al t_{\al\be}=(1+{b\over a})\d_\be t_\mu^\mu.\eqno(2.16)$$
Using $\d^2 v_\la=0$ we see that this condition is satisfied by
$t_{\al\be}=\d_\al v_\la\d_\be v^\la$ (2.14) with $a=2$, $b=-1$ 
as it must be, but
$t_{\al\be}$ is neither conserved nor traceless.

To be able to do non-perturbative calculations in massive gravity 
we look for the classical
theory corresponding to the coupling (2.13). It was shown in [1]
sect.5.5 that the pure graviton couplings $h\d h\d h$ correspond to the
Einstein-Hilbert Lagrangian
$$L_{\rm EH}={2\over\kappa^2}\sqrt{-g}R\eqno(2.17)$$
in the following sense. We write the metric tensor as
$$\sqrt{-g}g^{\mu\nu}=\eta^{\mu\nu}+\kappa h^{\mu\nu}\eqno(2.18)$$
and expand $L_{\rm EH}$ in powers of $\kappa$. Then the quadratic terms
$O(\kappa^0)$ give the free theory, the cubic terms $O(\kappa^1)$ agree
with the pure graviton coupling terms in (2.13) up to a factor 4 and so
on. The mass dependent terms are obtained if one introduces a
cosmological constant $-\Lambda=m^2/2$ in (2.17) [3], the minus sign
is due to the different convention in astrophysics [8]. As already said
we consider the massless limit in the following.

To obtain $T_v$ (2.14) we must add
$$L_v=\sqrt{-g}g^{\al\be}\d_\al v_\la\d_\be v^\la\eqno(2.19)$$
to $L_{\rm EH}$. One may be tempted to write covariant derivatives
$\nabla_\al$ instead of partial derivatives in order to get a true
scalar under general coordinate transformations. But this would produce
quartic couplings containing $v_\la$ and such terms are absent in the
quantum theory ([2], eq.(4.12)). For the same reason the Lorentz index
$\la$ in $v_\la$ is raised and lowered with the Minkowski tensor
$\eta_{\mu\nu}$, but all other indices with $g_{\mu\nu}$. Again this
might look odd to relativists, but the correct classical
expression is dictated by the quantum theory not by something else. Note
that the gauge variation (2.7) is just the infinitesimal form of general
coordinate transformations, so that this symmetry is built in on the 
quantum level not on the classical one.

The total classical Lagrangian we have to study is now given by
$$L_{\rm tot}=L_g+L_M+L_v\eqno(2.20)$$
where
$$L_g={c^3\over 16\pi G}\sqrt{-g}R\eqno(2.21)$$
we have introduced the velocity of light and Newton's constant for later
discussions. $L_M$ is given by the energy-momentum tensor $t_{\mu\nu}$ of
ordinary matter as usual. Strictly speaking we should also add a term
for the ghost coupling. At present we leave this out because probably
these unphysical degrees of freedom don't contribute in the classical
theory. 

Calculating first the variational derivative with respect to $g_{\mu\nu}$
we get the following modified Einstein equations
$$R_{\mu\nu}-{1\over 2}g_{\mu\nu}R-{8\pi G\over c^4}t_{\mu\nu}=
{16\pi G\over c^3}\Bigl(-\d_\mu v_\la\d_\nu v^\la+$$
$$+{1\over 2}g_{\mu\nu}g^{\al\be}\d_\al v_\la\d_\be v^\la\Bigl).\eqno(2.22)$$
Secondly the variation with respect to $v_\la$ gives
$$\d_\al(\sqrt{-g}g^{\al\be}\d_\be v^\la)=0.\eqno(2.23)$$
After multiplication with $1/\sqrt{-g}$ this is the Laplace-Beltrami or
rather the wave equation in the metric $g^{\al\be}$. The coupled system
(2.22) (2.23) is the modification of general relativity which we want to
study. 

\section{Modified Schwarzschild solution}

We wish to construct a static spherical symmetric solution of the
modified field equations. Following the convention of [5] we write the
metric as
$$ds^2=e^\nu c^2dt^2-r^2(d\te^2+\sin^2\te\d\fii^2)-e^\la dr^2\eqno(3.1)$$
where $\nu$ and $\la$ are functions of $r$ only. We take the coordinates
$x^0=ct$, $x^1=r$, $x^2=\te$, $x^3=\fii$ such that
$$g_{00}=e^\nu,\quad g_{11}=-e^\la$$
$$g_{22}=-r^2,\quad g_{33}=-r^2\sin^2\te\eqno(3.2)$$
and zero otherwise. The components with upper indices are the inverse of
this. The determinant comes out to be
$$g={\rm det}g_{\mu\nu}=-e^{\nu+\la}r^4\sin^2\te.\eqno(3.3)$$ 
Next we compute the Christoffel symbols
$$\Gamma^0_{00}={1\over 2}\nu',\quad \Gamma^1_{00}={1\over 2}\nu'e^{\nu-\la},
\quad \Gamma^1_{11}={1\over 2}\la',\quad \Gamma^1_{22}=-re^{-\la}$$
$$\Gamma^1_{33}=-re^{-\la}\sin^2\te,\quad \Gamma^2_{12}={1\over r},
\quad \Gamma^2_{33}=-\sin\te\cos\te,\quad \Gamma^3_{13}={1\over r},
\quad \Gamma^3_{23}=\cot\te$$
and zero otherwise, the prime denotes the derivative with respect to $r$
always. From this the contracted curvature tensor is given
by
$$R_{\mu\nu}={\d\Gamma^\la_{\mu\nu}\over\d x^\la}-{\d\Gamma^\la_{\mu\la} 
\over\d x^\nu}+\Gamma^\la_{\mu\nu}\Gamma^\ro_{\la\ro}-\Gamma^\ro_{\mu\la}
\Gamma^\la_{\nu\ro}.$$
We obtain
$$R_{00}={1\over 2}e^{\nu-\la}(\nu''+{1\over 2}\nu^{\prime 2}-{1\over 2}
\nu'\la'+{2\over r}\nu')\eqno(3.4)$$
$$R_{11}=-{1\over 2}\nu''+{1\over 4}\la'(\nu'+{4\over r})-{1\over
4}\nu^{\prime 2}\eqno(3.5)$$
$$R_{22}=e^{-\la}({r\over 2}\la'-{r\over 2}\nu'-1)+1\eqno(3.6)$$
$$R_{33}=\sin^2\te R_{22}.\eqno(3.7)$$
For the scalar curvature we then find
$$R=e^{-\la}\Bigl(\nu''+{1\over 2}\nu^{\prime 2}-{1\over 2}\nu'\la'+
2{\nu'\over r}-2{\la'\over r}+{2\over r^2}\Bigl)-{2\over r^2}.
\eqno(3.8)$$

Now we are ready to calculate
$$G_\mu^{\>\nu}\=d R_\mu^{\>\nu}-{1\over 2}\delta_\mu^\nu R,
\eqno(3.9)$$
$$G_0^{\>0}=e^{-\la}\Bigl({\la'\over r}-{1\over r^2}\Bigl)+{1\over r^2} 
\eqno(3.10)$$
$$G_1^{\>1}=-e^{-\la}\Bigl({\nu'\over r}+{1\over r^2}\Bigl)+{1\over r^2} 
\eqno(3.11)$$
$$G_2^{\>2}=G_3^{\>3}=-e^{-\la}\Bigl({\nu''\over r}+{\nu^{\prime 2} 
\over 4}-{\nu'\la'\over 4}-{\la'\over 2r}+{\nu'\over 2r}\Bigl). 
\eqno(3.12)$$
The energy momentum tensor of ordinary matter is assumed in the diagonal
form
$$t_\al^{\>\be}={\rm diag}(qc^2,-p_1,-p_2,-p_3)\eqno(3.13)$$
where $q(r)$ is the (inertial) mass density and $p_j(r)$ the presure components.

After collecting these well known results [5] we turn to the new
contributions involving the vector-graviton field $v^\kappa$. 
Since
$$\d_\al(\sqrt{-g}g^{\al\be}\d_\be)=-\sin\te{\d\over\d r}\Bigl(
e^{(\nu-\la)/2}r^2{\d\over\d r}\Bigl)-e^{(\nu+\la)/2}{\d\over\d\te}
\Bigl(\sin\te{\d\over\d\te}\Bigl)-$$
$$-{e^{(\nu+\la)/2}\over\sin\te}{\d^2\over\d\fii^2},\eqno(3.14)$$
the Laplace-Beltrami equation (2.23) can be written in the form
$${\d\over\d r}\Bigl(e^{(\nu-\la)/2}r^2{\d\over\d r}\Bigl)v_\kappa=
e^{(\nu+\la)/2}L^2v_\kappa,\eqno(3.15)$$
where $L^2$ is the quantum mechanical angular momentum operator squared.
Consequently the angular dependence of $v_\kappa$ is given by spherical
harmonics $Y_l^m(\te,\fii)$:
$$v_\kappa(r,\te,\fii)=v_\kappa(r)Y_l^m(\te,\fii)$$
and the radial function $v_\kappa(r)$ satisfies the following radial
equation
$${\d\over\d r}\Bigl(e^{(\nu-\la)/2}r^2{\d v_\kappa(r)\over\d r}\Bigl)=
l(l+1)e^{(\nu+\la)/2}v_\kappa(r).\eqno(3.16)$$
For what follows we consider the case $l=0$ where we get the simple result
$$v'_\kappa(r)={C_\kappa\over r^2}e^{(\la-\nu)/2},\eqno(3.17)$$
here $C_\kappa, \kappa=0,1,2,3$ are constants of integration.

Now we are ready to write down the modified radial Einstein equations.
In (2.22) we raise the second index
$$G_\mu^{\>\nu}={8\pi G\over c^3}\Bigl({1\over c}t_\mu^{\>\nu}+
\delta_\mu^\nu g^{\al\be}\d_\al v_\la\d_\be v^\la-2\d_\mu v_\kappa
\d^\nu v^\kappa\Bigl).\eqno(3.18)$$
$$G_0^{\>0}=e^{-\la}\Bigl({\la'\over r}-{1\over r^2}\Bigl)+{1\over r^2}
={8\pi G\over c^3}(q(r)c+w_0(r))\eqno(3.19)$$
$$-G_1^{\>1}=e^{-\la}\Bigl({\nu'\over r}+{1\over r^2}\Bigl)-{1\over r^2}
={8\pi G\over c^3}({p_1(r)\over c}+w_1(r))\eqno(3.20)$$
$$-G_2^{\>2}=e^{-\la}\Bigl({\nu''\over 2}+{\nu^{\prime 2}\over 4}- 
{\nu'\la'\over 4}-{\la'\over 2r}+{\nu'\over 2r}\Bigl)
={8\pi G\over c^3}({p_2(r)\over c}+w_2(r)),\eqno(3.21)$$
$$-G_3^{\>3}=e^{-\la}\Bigl({\nu''\over 2}+{\nu^{\prime 2}\over 4}- 
{\nu'\la'\over 4}-{\la'\over 2r}+{\nu'\over 2r}\Bigl)
={8\pi G\over c^3}({p_3(r)\over c}+w_3(r)),\eqno(3.22)$$
where $w_0, w_1, w_2, w_3$ are the contributions from the $v$-field
in (3.18). The two equations (3.19) (3.20) together with (3.15) are
sufficient to calculate all unknown functions $\la, \nu, v_\kappa$.
That means the equations (3.21) (3.22) are additional constraints.
It is a delicate problem to satisfy them. If we assume $p_2=p_3$, we must
have $w_2=w_3$ and this is only possible for $l=0$. Then $v_\kappa$ depends
on $r$ only. We restrict ourselves to this case in the following,i.e. $p_j=p$.

From (3.17) we now obtain the following contributions of the $v$-field
$$g^{11}\d_1v_\kappa\d_1v^\kappa={1\over r^4}e^{-\nu}(C_1^2+C_2^2+C_3^2
-C_0^2)$$
$$\=d {C\over r^4}e^{-\nu}\eqno(3.23)$$
and zero otherwise, so that
$$w_0={C\over r^4}e^{-\nu}=w_2=w_3\eqno(3.24)$$
$$w_1=-w_0.$$
We see in (3.19) that the mass density $q(r)$ gets enlarged by the quantity
$w_0(r)/c$ which we shall call dark density. We omit the notion ``matter''
because this energy density comes from the vector-graviton field, it is a relic
of the massive graviton.

Next by suitable combination we simplify the equations. Adding (3.19) to
(3.20) we get
$${e^{-\la}\over r}(\la'+\nu')=g(q+{p\over c^2})\eqno(3.25)$$
where
$$g={8\pi G\over c^2}.\eqno(3.26)$$
Eliminating $p$ from (3.20) and (3.21) we obtain
$$\nu''=2e^\la\Bigl(-{1\over r^2}+2g{w_0\over c}\Bigl)+{\nu'\la'\over 2}
-{\nu^{\prime 2}\over 2}+{\la'+\nu'\over r}+{2\over r^2}.
\eqno(3.27)$$
Next we differentiate (3.20) with respect to $r$ and use (3.27):
$$g\Bigl({p'\over c^2}-{w'_0\over c}\Bigl)=-e^{-\la}{\nu'\over 2r}
(\la'+\nu')+4{g\over cr}w_0.\eqno(3.28)$$
Substituting (3.25) inhere we finally arrive at
$${p'\over c^2}=-{\nu'\over 2}\Bigl(q+{p\over c^2}+{2\over c}w_0
\Bigl)\eqno(3.29)$$
where (3.24) has been used. This differential equation for the pressure
will be used instead of the second order equation (3.21) in the
following.

For a first orientation we want to solve the equations neglecting the
ordinarary matter, i.e. $q=0=p$. Then we find from (3.25) $\la'+\nu'=0$
or
$$\nu(r)=D_1-\la(r)\eqno(3.30)$$
where $D_1$ is a constant of integration. Using this is (3.24) we have
$$w_0={D_2\over r^4}e^\la.\eqno(3.31)$$
Multiplying (3.19) by
$$y=re^{-\la}\eqno(3.32)$$
we write the equation in the form
$$y'=1-8\pi{D_3\over ry}\eqno(3.33)$$
where
$$D_3={GD_2\over c^3}.\eqno(3.34)$$
Although this equation (3.33) looks rather simple, we did not succeed in
expressing the solution by known special funtions, but a numerical solution
can be easily obtained. The dark density $w_0$ (3.31) is given in terms of
$y$ (3.32) by
$$w_0={D_2\over r^3y(r)}.\eqno(3.35)$$

For large $r$ we find the following power series expansion
$$y=r-a_1+{a_2\over r}+{a_3\over r^2}+\ldots\eqno(3.36)$$
where
$$a_2=8\pi D_3,\quad a_3=4\pi D_3a_1,\ldots\eqno(3.37)$$
$a_1$ is related to the total mass $M$ by the Schwarzschild relation
$$a_1={2GM\over c^2}.\eqno(3.38)$$
This can be taken as the start of the numerical integration of (3.33)
from large to small $r$. Note that the Schwarzschild horizon disappears,
but there remains a singularity at $r=0$. It is of the form
$$y=4\sqrt{-\pi D_3\log r+\alpha+{\beta\over{\log r}}}.$$
This is too weak to render $r^2w_0$ (3.35) integrable at $r=0$. We must
conclude that the classical theory breaks down at $r=0$. This is a
phenomenon known from ordinary Einstein equations.

Without better options we cut off the singularity by assuming $w_0=
{\rm const.}$ inside a core radius $r_c$, for $r\ge r_c$ $w_0$ is given
by (3.35). The core radius $r_c$ is determined as follows: We compute
the total mass
$$M={4\pi\over c} \Bigl(\int\limits_0^{r_c}+\int\limits_{r_c}^\infty
\Bigl)r^2w_0(r)$$
$$={4\pi\over c}\Bigl({r_c^3\over 3}w_0(r_c)+D_2\int\limits_{r_c}^\infty
{dr\over ry}\Bigl)$$
$$={4\pi\over c}\Bigl({D_2\over 3y(r_c)}+{D_2\over 8\pi D_3}\int
\limits_{r_c}^\infty(1-y')dr\Bigl)$$
$$={4\pi\over 3c}{D_2\over y(r_c)}-{c^2\over 2G}(-a_1+r_c-y(r_c))$$
where (3.34) has been used. By (3.38) $M$ drops out and we get $r_c$ as
the zero of the equation
$$r_c-y(r_c)={8\pi\over 3}{D_3\over y(r_c)}.\eqno(3.39)$$
Varying $D_3$ this enables us to vary $r_c$ and the radius of the halo,
if we keep its total mass $M$, i.e. $a_1$ (3.38) fixed. The differential
equation (3.33) is invariant under the scale transformation
$$r\to\alpha r,\quad y\to\alpha y,\quad D_3\to\alpha^2 D_3,\quad
a_1\to\alpha a_1.\eqno(3.40)$$
This implies that the length scale of the halo can be freely adjusted.

In astrophysical applications we measure $r$ in kpc. Let us consider a
halo of $10^{14}$ solar masses, then $a_1$ (3.38) is equal to $10^{-2}$
(in kpc). Taking $D_3$ of the order $10^{-3}$ we find rather small core
radii of a few kpc. Precisely, our numerical results agree with the linear
relation
$$r_c=3352 D_3.\eqno(3.41)$$
Outside $r_c$, $y$ is given by the four terms in (3.36) up to promille
accuracy. That means the dark density profile (3.35) can be approximated by
$$w_0={D_2\over r}{1\over r^3-a_1r^2+a_2r+a_3},\quad r\ge r_c.\eqno(3.42)$$
Then the dark mass interior to $r$ is given by
$$M(r)=4\pi\int\limits^r w_0r^2dr=4\pi D_2\int\limits^r{r\,dr\over
r^3-a_1r^2+a_2r+a_3}.\eqno(3.43)$$
Expanding this for small $r$ (but $>r_c$) we find a linear dependence
$$M(r)=\alpha_0+\alpha_1r+\ldots\eqno(3.44)$$
This corresponds to constant orbital velocity as observed in dark halos
of spiral galaxies.

On the other hand Navarro, Frenk and White [6] have found the following profile
$$\ro(r)={\ro_s\over r(1+{r\over r_s})^2}$$
from N-body simulations, where $r_s$ is the so-called scaling radius and $\ro_s$ 
the scaling density. In contrast to our result (3.42), $\ro (r)$ only decreases
as $1/r^3$ for large $r$. But then $r^2\ro (r)$ is not integrable
at infinity. A profile with $1/r^4$ tail was proposed by Hernquist [7] for
elliptical galaxies. It remains to be seen which profile gives the best
agreement with reality.

{\bf Acknowledgement.} Illuminating discussions with Dan Grigore, Klaus Fredenhagen,
Andreas Aste and computational aid by Roland Bernet are gratefully acknowledged.


\begin{thebibliography}{} 

\bibitem{} G. Scharf, Quantum Gauge Theories - A True Ghost Story,
Wiley-Interscience, New York 2001

\bibitem{} D.R. Grigore, G. Scharf, Gen.Relativ.Grav. {\bf 37} (6) 
1075 (2005), hep-th/0404157 

\bibitem{} D.R. Grigore, G. Scharf, Massive gravity from descent 
equations, hep-th/0711.0869

\bibitem{} J.B. Berchtold, G. Scharf, Gen.Relativ.Grav. {\bf 39 (9)} 
(2007) 1489, hep-th/0702181

\bibitem{} R. Adler, M. Bazin, M. Schiffer, Introduction to general
relativity, McGraw-Hill, New York 1965

\bibitem{} J.F. Navarro, C.S. Frenk, S.D.M. White, Astrophys.Journ.
{\bf 462} 563 (1996)

\bibitem{} L. Hernquist, Astrophys.Journ. {\bf 356} 359 (1990)

\bibitem{} E.W. Kolb, M.S. Turner, The Early Universe, Addison-Wesley
Publishing Company, Redwood City 1990

\end{thebibliography}
\end{document}